\documentclass{egpubl}
\usepackage{eurovis2023}
\usepackage{amsmath}
\usepackage{amsfonts}

\EuroVisShort  

\usepackage[T1]{fontenc}
\usepackage{dfadobe}  

\usepackage{cite}  
\BibtexOrBiblatex
\electronicVersion
\PrintedOrElectronic
\ifpdf \usepackage[pdftex]{graphicx} \pdfcompresslevel=9
\else \usepackage[dvips]{graphicx} \fi

\usepackage{egweblnk}
\copyrightTextTitPag{}
\copyrightTextRunPag{}
\makeatletter
\renewcommand{\ps@titlepage}{%
  \let\@mkboth\@gobbletwo
  \let\@oddhead\@empty
  \let\@evenhead\@empty
  \let\@oddfoot\@empty
  \let\@evenfoot\@empty
}
\let\@oddfoot\@empty
\let\@evenfoot\@empty
\makeatother
\hypersetup{pdfsubject={},pdfkeywords={}}

\title[Neural Caching of Prefiltered Radiance for Specular Lighting]%
      {Neural Caching of Prefiltered Radiance for Specular Lighting \vspace{-3ex}}

\author[D. Klepikov]
{\parbox{\textwidth}{\centering D.\, Klepikov\orcid{0000-0001-9577-7047}, V.\, Frolov\orcid{0000-0001-8829-9884}
}
        \\
{\parbox{\textwidth}{\centering Lomonosow Moscow State University}
}
}

\begin{document}

\teaser{
  \includegraphics[width=\linewidth]{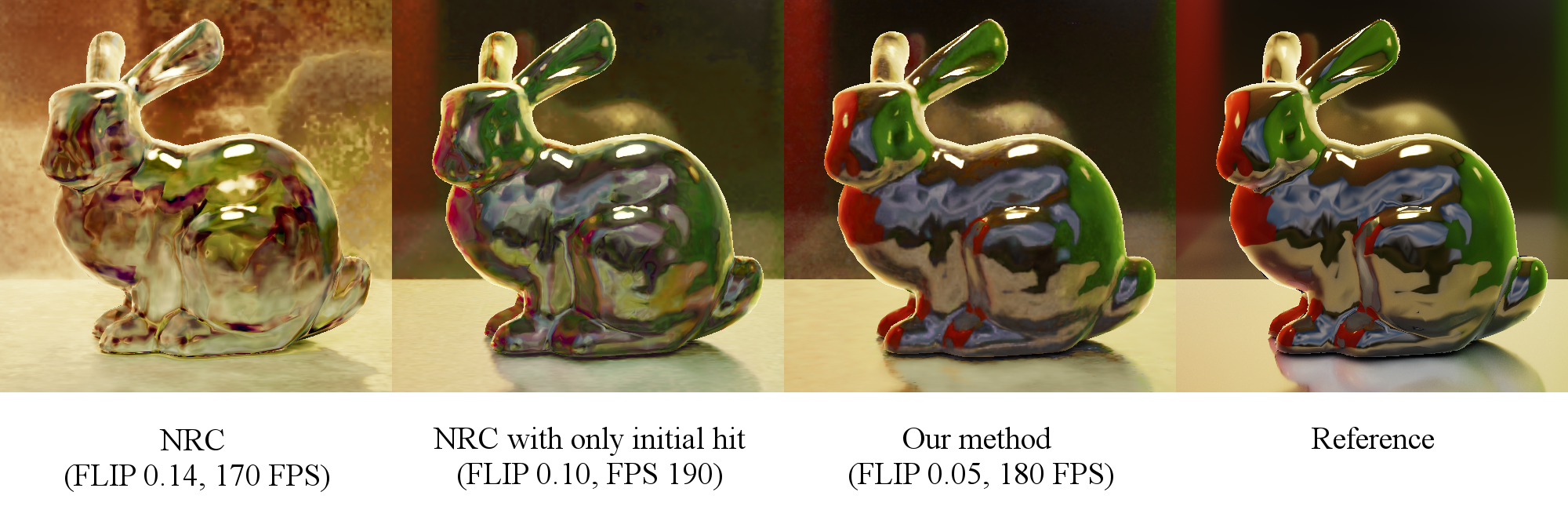}
  \includegraphics[width=\linewidth]{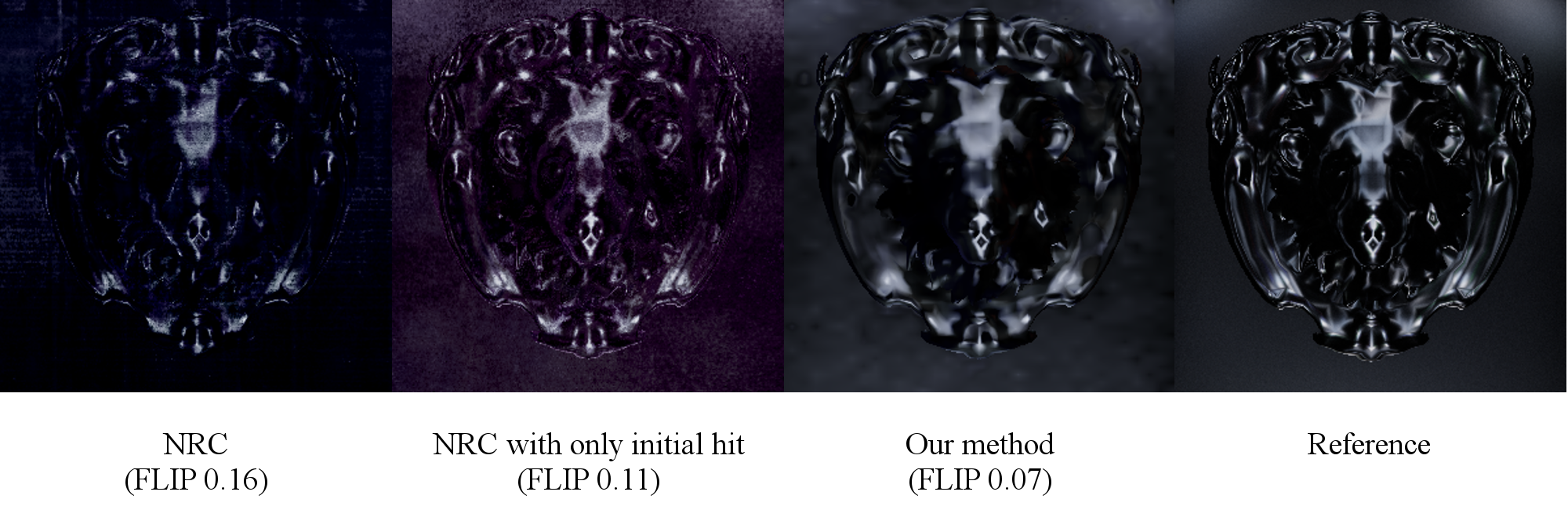}
  \centering
   \caption{
        Qualitative comparison on Bunny (first row) and Specular Sponza (second row), a version of Sponza with highly reflective, untextured surfaces. All renderings use only specular lobes. In Bunny, the walls and floor have roughness 0.1. The NRC baseline uses the implementation provided in~\cite{muller2021real}. The reference is obtained by accumulating the Falcor path tracer over 4000 frames.}
 \label{fig:teaser}
}

\maketitle

\begin{abstract}
Neural Radiance Caching (NRC)~\cite{muller2021real} provides an online neural representation of scene illumination for real-time path tracing. This paper presents an NRC variant tailored to specular lighting through a reflection-direction parameterization and a roughness-dependent radiance target. Our network predicts prefiltered incoming radiance at individual surface points, which is combined with a precomputed BRDF integration map using the split-sum approximation to estimate outgoing radiance. On Bunny and Specular Sponza, the reported comparisons indicate faster convergence to the accumulated path-tracing reference and improved specular-lighting quality relative to the evaluated NRC baselines. Our online-trained method maintains real-time performance while adapting its cache during rendering.


\end{abstract}  
\section{Introduction}

Specular reflection depends strongly on viewing direction and surface roughness, making its representation challenging for radiance caching. Neural Radiance Caching (NRC)~\cite{muller2021real} learns scene illumination online to accelerate path tracing. We target specular lighting by parameterizing the cache with the reflected viewing direction and learning roughness-dependent prefiltered incoming radiance for split-sum reconstruction.

The main goal of path-tracing is to calculate the solution of Light Transport Equation (LTE) \cite{LTE}:
\begin{equation} 
L_o(\omega_o) = L_e(\omega_o) + \int_{H^+} L_i(\omega_i) f_s(\omega_i,\omega_o) \cos(\theta_i)\,d\omega_i.
\label{LTE_expr} 
\end{equation} 
Here, \(L_e\) is the radiance emitted in direction \(\omega_o\), \(L_i\) is the radiance arriving from direction \(\omega_i\), and \(f_s\) is the bidirectional scattering distribution function (BSDF). The angle \(\theta_i\) is measured between the surface normal \(n\) at position \(x\) and \(\omega_i\). We omit the position argument for brevity. For reflection, the BSDF is restricted to a bidirectional reflectance distribution function (BRDF). In this work, we consider its specular component. Estimating the integral is computationally expensive, motivating approaches that cache previously computed lighting information.

Traditional radiance caching approaches like\cite{krivanek2006making} (and analogues) are based on several heuristics that are difficult to implement and tend to produce artefacts that need special care to be suppressed \cite{schwarzhaupt2012practical}. Other progressive rendering and reconstruction approaches (for example \cite{moon2014adaptive}) also uses handcrafted mathematic models which makes their implementation and testing inside and existing rendering system a challenging task. At the same time NRC\cite{muller2021real} reduces such complexity by way of abstraction of internal radiance cache representation via neural network. This method represents scene illumination with a multilayer perceptron that maps hit ray positions and directions to radiance values of the scattered light. The approach is lightweight and it is designed to continuously train in real-time to support dynamic scenes while ensuring predictable performance and resource usage. The research presents a multitude of methods to regulate and stabilize the output of the neural network like self-training and path termination that are omitted in our paper. The idea of encoding input data was also explored in these papers \cite{muller2022instant} \cite{verbin2022ref} \cite{muller2019neural}. 

Our method parameterizes the cache to represent specular reflection, including mirror-like and rough glossy reflection. Following the reflection-direction parameterization used in~\cite{verbin2022ref}, we use the viewing direction reflected about the surface normal as a network input. We also use the split-sum approximation~\cite{karis2013real} to separate prefiltered lighting from the final BRDF reconstruction. The cache predicts roughness-dependent filtered radiance, while the BRDF integration lookup supplies the final Fresnel-dependent reconstruction. Together, these terms approximate the reflected component of Equation~\eqref{LTE_expr}.

\begin{figure*}[t!]
  \centering
  \mbox{} \hfill
  \includegraphics[width=\linewidth]{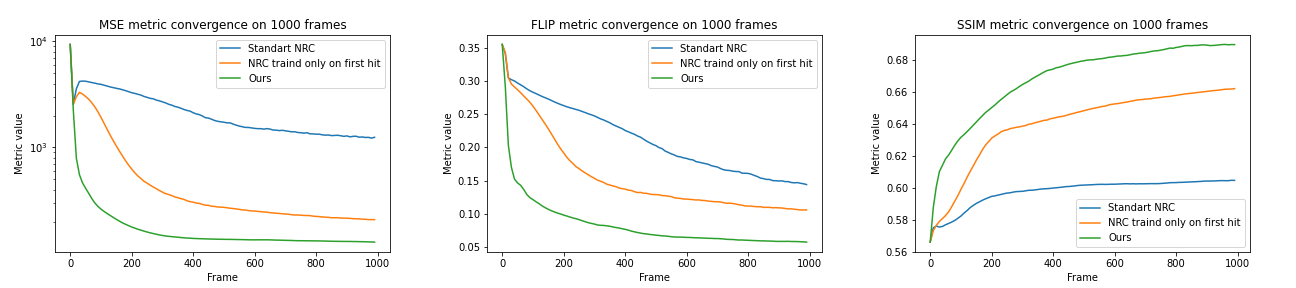}
  \hfill
  \includegraphics[width=\linewidth]{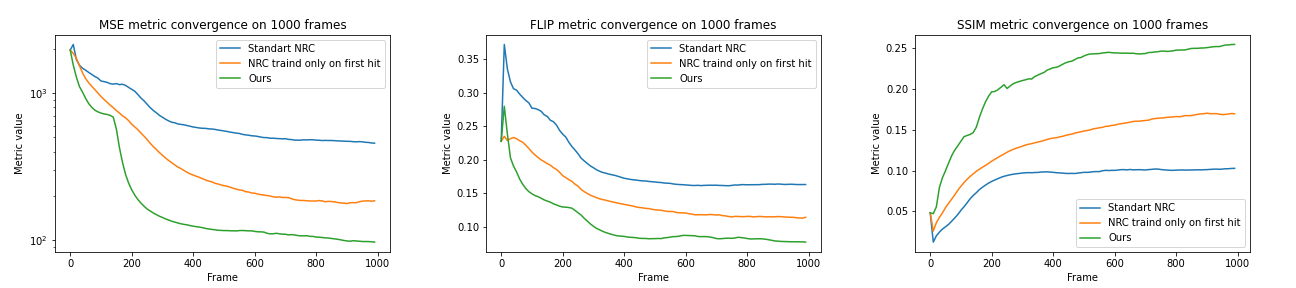}
  \hfill \mbox{}
  \caption{\label{fig:ex2}%
        FLIP, SSIM, and MSE during online training on Bunny (first row) and Specular Sponza (second row). The horizontal axis measures rendering frames. Lower FLIP and MSE values and higher SSIM values indicate better agreement with the reference. MSE is displayed on a logarithmic scale. The curves compare our method with the evaluated NRC variants. }
\end{figure*}

\section{Previous Work}

\subsection{Microfacet Specular BRDF}

Microfacet theory is a popular approach to modelling surface reflection and transmission. In this theory, surfaces are modelled as a collection of small microfacets, with their aggregated statistical behaviour determining the observed scattering of light. The microfacet specular BRDF is an expression that describes the scattering of light from a surface composed of microfacets. To compute BRDF from such a model, local lighting effects at the microfacet level need to be considered, such as microfacet occlusion, shadowing, and interreflection. An expression representing BRDF consists of two statistical measures, a microfacet normalized distribution function(NDF) of facets represented as \(D\) and a shadowing-masking function between facets \(G\), and a Fresnel reflectance factor \(F\). All of that is combined in a general Cook-Torrance \cite{cook1981reflectance}\cite{cook1982reflectance} microfacet specular shading model. Here \(h\) is the halfway-vector between \(\omega_{i},\omega_{o}\).
\begin{equation} 
f_{r}(\omega_{i},\omega_{o}) = \frac{D(h)F(\omega_{o},h)G(\omega_{i},\omega_{o})}{4(n \cdot \omega_{i})(n \cdot \omega_{o})}  
\label{cooktorr} 
\end{equation} 
For the implementation of those terms, we used ones implemented in the Falcor rendering framework \cite{Falcor} and thoroughly described in Disney’s research paper \cite{burley2012physically}. In the presented paper NDF is represented as a modified Trowbridge-Reitz(GGX)\cite{trowbridge1975average} distribution. For the Fresnel term the Schlick approximation \cite{schlick1994inexpensive} is used. For the shadowing-masking function modified GGX model\cite{walter2007microfacet} is used with reparametrization presented in Disney’s research paper to dampen the shine on small roughness values.

\subsection{Split sum approximation}

The reflected component of Equation~\eqref{LTE_expr} can be estimated by importance sampling. Let \(p(\omega_i)\) denote the probability density of the sampled incoming direction, measured with respect to solid angle. We use the NDF-based sampling implementation provided by Falcor. For \(N\) samples \(\omega_{i,k}\), the estimator is

\begin{multline} 
L_o(\omega_o) = L_e(\omega_o) + \int_{H^+} L_i(\omega_i)f_r(\omega_i,\omega_o)\cos(\theta_i)\,d\omega_i\\
\approx L_e(\omega_o) + \frac{1}{N}\sum_{k=1}^{N}\frac{L_i(\omega_{i,k})f_r(\omega_{i,k},\omega_o)\cos(\theta_{i,k})}{p(\omega_{i,k})}.
\label{splitsumpre} 
\end{multline} 

Even with importance sampling, many samples may be required for high-quality results. Karis~\cite{karis2013real} approximates the reflected-radiance estimate by separating it into two averages:

\begin{multline} 
L_o(\omega_o)-L_e(\omega_o)\\
\approx \left(\frac{1}{N}\sum_{k=1}^{N}L_i(\omega_{i,k})\right)
\left(\frac{1}{N}\sum_{k=1}^{N}\frac{f_r(\omega_{i,k},\omega_o)\cos(\theta_{i,k})}{p(\omega_{i,k})}\right).
\label{splitsum} 
\end{multline} 

The first factor represents prefiltered environment radiance. The practical prefilter in~\cite{karis2013real} samples the GGX normal distribution and uses normalized cosine weighting. During prefiltering, the viewing direction is assumed perpendicular to the filtering surface, so \(n=\omega_o\). The filtering normal is aligned with the reflected viewing direction used for lookup. Incoming directions \(\omega_i\) remain sampled over the hemisphere. This normal-view assumption introduces approximation error at grazing viewing angles.

The second factor is the BRDF integration map. With Schlick's Fresnel approximation, it is represented by two coefficients stored in a lookup texture indexed by roughness and the cosine of the viewing angle. Our method uses this split-sum factorization and BRDF integration lookup to reconstruct outgoing radiance from the neural cache.

\subsection{Neural radiance cache}

We examined various methods employing neural networks (and multilayer perceptron specifically) for providing scene information. Neural methods are increasingly being used for rendering in various applications. The recent method of spatio-directional neural representations replaces traditional discrete representations with an MLP for reconstructing 3D geometry-aware solutions. The most widely adopted strategy currently is Neural Radiance Fields\cite{mildenhall2020nerf}, the science publication of which inspired a multitude of works \cite{barron2021mip} \cite{verbin2022ref} in the novel view synthesis problem. 

Gradually, the interest in neural approaches for global illumination increased. There has been a significant effort to improve sampling\cite{muller2019neural} and denoising of Monte Carlo global illumination algorithms \cite{bako2017kernel} \cite{chaitanya2017interactive} \cite{gharbi2019sample}. The current work is based on NRC \cite{muller2021real} where researchers use small neural networks that undergo online training to store and predict indirect global illumination for real-time path tracing. Further development shows the significant advantage of using hash grid parametric encoding for NRC\cite{muller2022instant}.


\section{Proposed method}

Our method uses a single multilayer perceptron to predict prefiltered incoming radiance at a queried position, reflection direction, and roughness. The network is trained during rendering to adapt to scene changes. We divide rendering into three passes. The first generates training paths with multiple surface bounces. The second traces rendering paths and stores the surface information needed to query the cache, terminating paths earlier than the training pass. Both passes use Monte Carlo integration and importance sampling for continuation directions. The third queries the cache and reconstructs outgoing radiance using the BRDF integration lookup.

For each surface hit, we collect the radiance and path-throughput information along the continuation of the training path. These data are used to reconstruct an RGB incoming-radiance target \(Y\), which is stored in the training buffer together with its cache query. Different sampled paths can produce different targets for the same query.

The cache query is \(q=(\phi(x),\omega_{\mathrm{refl}},\rho)\), where \(x\) is the surface position, \(\phi\) is its hash-grid encoding, and \(\rho\) is the surface roughness. For an outgoing direction \(\omega_o\) pointing away from the surface, the reflected direction is \(\omega_{\mathrm{refl}}=2(n\cdot\omega_o)n-\omega_o\). We hash-grid encode positions~\cite{muller2022instant} and pass reflection direction and roughness directly to the MLP. Each query produces an RGB estimate of the filtered incoming radiance. The roughness input conditions the cached lighting on the reflection lobe.

We train the network using mean squared error (MSE). Under squared-error loss, the population-optimal predictor is the conditional mean of the training target given the cache query,
\begin{equation}
\widehat{L}^{\,*}(q)=\mathbb{E}[Y\mid q].
\label{conditional_mean}
\end{equation}
The sampling distribution and target weighting therefore determine the radiance filter represented by the cache. The network and optimization configuration follow~\cite{muller2021real}. To reduce fluctuations during online adaptation, we use an exponential moving average of the optimized network weights. The final pass combines the predicted radiance with the stored BRDF integration map to estimate outgoing specular radiance in the camera direction. This reconstruction retains the split-sum approximation and its grazing-angle limitations.


\section{Experiments and Results}

We use the ray-tracing pipeline in Falcor~\cite{Falcor} and the MLP implementation in tiny-cuda-nn~\cite{tiny-cuda-nn}. All images were rendered at a resolution of \(1920\times1080\) on an Intel Core i7-8700K processor with an NVIDIA RTX 3080 Ti graphics card. We evaluate image quality using FLIP~\cite{andersson2020flip}, SSIM~\cite{wang2004image}, and MSE on each rendering frame. MSE is displayed on a logarithmic scale because its values span a wide range. The reference image is obtained by accumulating the regular path tracer over 4000 frames.

Our findings indicate that the modifications have not resulted in any discernible performance degradation while improving convergence and output quality. The BRDF integration map is precomputed and queried during rendering. The changes to the neural network inputs do not negatively influence the other major NRC components.

We compare our method with the NRC baseline and an NRC variant trained only on initial hits. We include the initial-hit variant because removing training targets from multiple hits along a path reduced NRC output quality in our tests. The two variants are distinguished in the comparison plots.

Figure~\ref{fig:teaser} presents a qualitative comparison after 1000 rendering frames. The displayed images were adjusted for improved visibility. The FLIP results indicate closer agreement with the accumulated reference for our method than for the evaluated NRC baselines. Figure~\ref{fig:ex2} shows the evolution of the quality metrics during training and indicates faster convergence for our method. On Bunny, our method achieves better image quality than the NRC baseline after 1000 frames.

\section{Conclusion}

We presented an NRC variant for real-time specular-lighting approximation that predicts roughness-dependent prefiltered incoming radiance and reconstructs outgoing radiance using the split-sum approximation. On Bunny and Specular Sponza, the reported comparisons show faster convergence and improved output quality relative to the evaluated NRC baselines while maintaining the same level of performance. The evaluated method uses hash-grid position encoding, a reflection-direction input, and split-sum reconstruction. It retains the grazing-angle limitations of the split-sum approximation. Online training enables the cache to adapt during rendering, motivating further investigation of its behavior in dynamic scenes.


\bibliographystyle{eg-alpha-doi}
\bibliography{egbibsample}


\newpage

\end{document}